\documentclass[english,prd,twocolumn,superscriptaddress,nofootinbib,preprintnumbers]{revtex4}
\usepackage[latin1]{inputenc}
\usepackage{graphicx}
\usepackage{amssymb}

\usepackage{amsfonts}
\usepackage{dcolumn}
\usepackage{hyperref}

\def\be{\begin{equation}}
\def\ee{\end{equation}}
\def\ba{\begin{eqnarray}}
\def\ea{\end{eqnarray}}
\def\bs{\begin{subequations}}
\def\es{\end{subequations}}

\newcommand{\rd}{{\rm d}}

\usepackage{color}

\makeatother

\usepackage{babel}
\makeatother
\begin{document}

\title{Assisted dark energy}

\author{Junko Ohashi}
\affiliation{Department of Physics, Faculty of Science, Tokyo University of Science, 
1-3, Kagurazaka, Shinjuku-ku, Tokyo 162-8601, Japan}

\author{Shinji Tsujikawa}
\affiliation{Department of Physics, Faculty of Science, Tokyo University of Science, 
1-3, Kagurazaka, Shinjuku-ku, Tokyo 162-8601, Japan}

\begin{abstract}

Cosmological scaling solutions, which give rise to a scalar-field density 
proportional to a background fluid density during radiation and matter eras, 
are attractive to alleviate the energy scale problem of dark energy.
In the presence of multiple scalar fields the scaling solution can exit to 
the epoch of cosmic acceleration through the so-called assisted inflation 
mechanism. We study cosmological dynamics of 
a multi-field system in details with a general Lagrangian density 
$p=\sum_{i=1}^n X_i g(X_ie^{\lambda_i \phi_i})$, where 
$X_i=-(\nabla \phi_i)^2/2$ is the kinetic energy of 
the $i$-th field $\phi_i$, $\lambda_i$ is a constant, 
and $g$ is an arbitrary function in terms 
of $Y_i=X_ie^{\lambda_i \phi_i}$.
This covers most of the scalar-field models of dark energy proposed in 
literature that possess scaling solutions.
Using the bound coming from Big-Bang-Nucleosynthesis and the condition 
under which the each field cannot drive inflation as a single component of
the universe, we find the following features: 
(i) a transient or eternal cosmic acceleration can be realized 
after the scaling matter era,  
(ii) a ``thawing'' property of assisting scalar fields is crucial to determine
the evolution of the field equation of state $w_{\phi}$, and 
(iii) the field equation of state today can be consistent with the
observational bound $w_{\phi}<-0.8$ in the presence of multiple
scalar fields.

\end{abstract}

\date{\today}

\maketitle

\section{Introduction}

The constantly accumulating observational data continue to confirm 
the existence of dark energy responsible for cosmic acceleration 
today \cite{review}.
The cosmological constant, whose equation of state is $w=-1$, has been 
favored by the combined data analysis of supernovae Ia \cite{SNIa}, 
cosmic microwave background \cite{CMB}, and 
baryon acoustic oscillations \cite{BAO}. 
Meanwhile, if the cosmological constant originates from a vacuum 
energy associated with particle physics, its energy scale is enormously 
larger than the observed value of dark energy 
($\rho_{\rm DE} \approx 10^{-47}\,{\rm GeV}^4$).
Hence it is important to pursue an alternative possibility to
construct dark energy models consistent with particle physics.

Scalar-field models such as quintessence \cite{quin1,quin2} and 
k-essence \cite{kes} have been proposed to alleviate the above
mentioned problem.
In general the energy density of a scalar field $\phi$ dynamically 
changes in time, so that its value around the beginning of 
the radiation-dominated epoch can be much larger than the 
dark energy density today.
One of such models is quintessence with an exponential potential 
$V(\phi)=V_0 e^{-\lambda \kappa \phi}$ \cite{Ferreira,CLW}, 
where $\lambda$ is a constant and $\kappa=\sqrt{8\pi G}$ with 
$G$ being gravitational constant (see Ref.~\cite{Halliwell} for 
the classification of cosmological dynamics and also Refs.~\cite{earlyexp}
for early related papers). 
In fact, in higher-dimensional gravitational theories such as superstring 
and Kaluza-Klein theories, exponential potentials often appear from 
the curvature of internal spaces associated with the geometry of 
extra dimensions (so called ``modulus'' fields) \cite{expmoti}.
Moreover it is known that exponential potentials can arise
in gaugino condensation as a non-perturbative effect \cite{gaugino}
and in the presence of supergravity corrections to 
global supersymmetric theories \cite{CNR}.

The quintessence with an exponential potential 
$V(\phi)=V_0 e^{-\lambda \kappa \phi}$
gives rise to two distinct fixed points in the 
flat Friedmann-Lema\^{i}tre-Robertson-Walker (FLRW) background \cite{CLW}: 
(a) the scaling solution, and 
(b) the scalar-field dominated solution. 
If the slope $\lambda$ of the potential satisfies the
condition $\lambda>\sqrt{3(1+w_f)}$, where $w_f$ is the 
equation of state of a background fluid, then the solutions approach 
the scaling attractor characterized by a field density parameter 
$\Omega_{\phi}=3(1+w_f)/\lambda^2$.
Even if the field energy density $\rho_{\phi}$ is initially comparable 
to the background fluid density $\rho_f$, the field eventually 
enters the scaling regime in which $\rho_{\phi}$ is proportional to 
$\rho_f$. This is attractive to alleviate the fine-tuning problem 
of the energy scale of dark energy. 
However the scaling solution needs to exit from the matter 
era to the epoch of a late-time cosmic acceleration.
The scalar-field dominated solution ($\Omega_{\phi}=1$) can be 
an accelerated attractor for $\lambda<\sqrt{2}$, but this is
incompatible with the condition $\lambda>\sqrt{3(1+w_f)}$
required for the existence of scaling solutions.
Hence the scaling solution cannot be followed by 
the scalar-field dominated solution responsible for dark energy.

There are a number of ways to allow a transition from the scaling regime
to the epoch of cosmic acceleration.
One of them is to introduce a single-field potential that becomes shallow at late times, 
e.g., $V(\phi)=c_1 e^{-\lambda \kappa \phi}+c_2 e^{-\mu \kappa \phi}$
with $\lambda>\sqrt{3(1+w_f)}$ and $\mu<\sqrt{2}$ \cite{Barreiro}
(see Ref.~\cite{Jarv} for the classification of dynamics 
and Refs.~\cite{otherexp} for related works).
For this double exponential potential the field equation of state $w_{\phi}$ 
of the final attractor is given by $w_{\phi}=-1+\mu^2/3$. 
In order to satisfy the observational constraint 
$w_{\phi} \lesssim -0.8$ \cite{obsercon} today, 
we require that $\mu$ is smaller than the order of 1.
If the exponential potential originates from particle physics models
then the slope $\mu$ is typically larger than 1, which is difficult to 
be compatible with the condition for cosmic acceleration.

Another way is to consider multiple scalar fields with exponential 
potentials, e.g., $V(\phi_1, \phi_2)=c_1 e^{-\lambda_1 \kappa \phi_1}
+c_2 e^{-\lambda_2 \kappa \phi_2}$ \cite{Coley,KLT}
(see also Refs.~\cite{phantom}).
In fact such potentials arise from the compactification of 
higher dimensional theories to 4-dimensional space-time.
It is known that the phenomenon called {\it assisted inflation} \cite{Liddle98}
occurs for the multi-field exponential potential, even if the individual field
has too steep a potential to lead to cosmic acceleration
(see also Refs.~\cite{assistedpapers}).
For the sum of steep potentials satisfying the condition 
$\lambda_i>\sqrt{2}$ ($i=1, 2, \cdots, n$),
the multiple fields evolve to give dynamics matching 
a single-field model with 
$\lambda_{\rm eff}=\left( \sum_{i=1}^n 1/\lambda_i^2 
\right)^{-1/2}<\sqrt{2}$ \cite{Liddle98}.
Since the conditions $\lambda_i>\sqrt{2}$ are mostly satisfied
for the models motivated by particle physics, this cooperative 
accelerated expansion is attractive for both inflation and dark energy.
If we apply this scenario to dark energy, the scaling radiation and matter 
eras can be followed by the epoch of assisted acceleration as more fields
join the scalar-field dominated attractor with an effective equation of 
state $w_{\phi}=-1+\lambda_{\rm eff}^2/3$ \cite{KLT}.

The scaling solution arises not only for quintessence with an exponential potential
but also for more general scalar-field models with the Lagrangian density $p(\phi, X)$, 
where $X=-g^{\mu \nu}\partial_{\mu}\phi \partial_{\nu} \phi \equiv
-(\nabla \phi)^2/2$ is a kinetic term of the field $\phi$.
Here $g^{\mu \nu}$ is a metric tensor with the notation $(-,+,+,+)$.
It was found in Refs.~\cite{PT04,TS04} that the existence of scaling solutions 
restricts the form of the Lagrangian density to be $p(\phi, X)=Xg (Xe^{\lambda \phi})$, 
where $\lambda$ is a constant and $g$ is an arbitrary function in terms 
of $Y=Xe^{\lambda \phi}$ (here we use the unit $\kappa^2=1$). 
The quintessence with an exponential potential ($p=X-ce^{-\lambda \phi}$)
corresponds to the choice $g=1-c/Y$, whereas the choice $g=-1+cY$
gives rise to the dilatonic ghost condensate model: 
$p=-X+ce^{\lambda \phi}X^2$ \cite{TS04} 
(which corresponds to the string-theory motivated 
generalization of the ghost condensate 
model proposed in Ref.~\cite{Arkani}).
The tachyon Lagrangian density $p=-V(\phi)\sqrt{1-2X}$ with 
$V(\phi) \propto \phi^{-2}$ \cite{Paddy} also follows from the above scaling 
Lagrangian by a suitable field redefinition \cite{CGST}.

For the multi-field scaling Lagrangian density
$p=\sum_{i=1}^n X_i g(X_ie^{\lambda_i \phi_i})$
it was shown in Ref.~\cite{Tsuji06} that assisted inflation 
occurs with the effective slope 
$\lambda_{\rm eff}=\left( \sum_{i=1}^n 1/\lambda_i^2 \right)^{-1/2}$, 
irrespective of the form of $g$.
Hence one can expect that the scaling solution is followed by 
the assisted acceleration phase for such a general Lagrangian.
If we consider loop or higher-order derivative corrections 
to the tree-level action motivated from string theory 
(such as $e^{\lambda \phi}X^2$), the constant $\lambda$ is 
typically of the order of unity \cite{stringreview}.
In the single-field case this is not compatible with the condition 
for cosmic acceleration.
It is of interest to see how the presence of multiple
fields changes this situation.

In this paper we shall study cosmological dynamics of 
multiple scalar fields with the Lagrangian density 
$p=\sum_{i=1}^n X_i\,g(X_ie^{\lambda_i \phi_i})$.
We are interested in the case where the scaling radiation 
and matter eras induced by a field $\phi_1$ are followed 
by the dark energy dominated epoch assisted by other 
scalar fields. 
For the two-field quintessence with exponential potentials
a similar analysis was partially done in Ref.~\cite{KLT}, 
but we shall carry out detailed analysis by taking into account bounds 
coming from Big-Bang-Nucleosynthesis (BBN) and supernovae
observations. In particular the evolution of the field equation of 
state $w_{\phi}$ will be clarified in the presence of 
two and more than two fields.
We also investigate cosmological dynamics for the multi-field 
dilatonic ghost condensate model as an example of k-essence models.

This paper is organized as follows.
In Sec.~\ref{mmodel} we present the dynamical equations for
our general multi-field Lagrangian density without specifying 
any form of $g$. In Sec.~\ref{fixed} we derive the fixed points
that correspond to the scaling radiation/matter solutions
and the assisted field-dominated attractor.
In Secs.~\ref{exponential} and \ref{dilatonic} we study 
the multi-field cosmological dynamics for quintessence 
with exponential potentials and the dilatonic 
ghost condensate model, respectively.
Sec.~\ref{conclusions} is devoted to conclusions.

\section{Dynamical system}
\label{mmodel}

Let us first briefly review single-field scaling models
with the Lagrangian density $p(\phi, X)$.
The existence of cosmological scaling solutions demands 
that the field energy density 
$\rho_\phi=2Xp_{,X}-p$, where $p_{,X} \equiv \partial p/\partial X$, 
is proportional to the background fluid density $\rho_f$.
Under this condition the Lagrangian density is restricted 
to take the following form in the flat FLRW background \cite{PT04, TS04}
\begin{equation}
p(\phi, X)=X\,g(Xe^{\lambda \phi})\,,
\label{plagform}
\end{equation}
where $\lambda$ is a constant and $g$ is an arbitrary function
in terms of $Y \equiv Xe^{\lambda \phi}$.
The Lagrangian density (\ref{plagform}) is valid even in the 
presence of a constant coupling $Q$ between the field $\phi$
and non-relativistic matter and also in the presence of 
a Gauss-Bonnet (GB) coupling between the field and the
GB term\footnote{It is also possible to obtain a generalized form 
of the scaling Lagrangian density even when the coupling $Q$
between $\phi$ and non-relativistic matter is 
field-dependent \cite{AQTW}.} \cite{ST06}.
In the following we do not take into account such couplings.
Throughout this paper we use the unit $\kappa^2=8\pi G=1$.

The field density parameter for scaling solutions 
is given by $\Omega_\phi=3(1+w_f)p_{,X}/\lambda^2$ \cite{Tsuji06,AQTW}, 
where $w_f$ is the fluid equation of state.
If the field enters the scaling regime during the radiation era, 
the BBN places the bound 
$\Omega_\phi<0.045$ at the $2\sigma$ confidence 
level \cite{Bean}. This then gives the constraint 
$\lambda^2/p_{,X}>88.9$.

Besides scaling solutions, there is a scalar-field dominated point ($\Omega_\phi=1$)
with the equation of state $w_{\phi}=-1+\lambda^2/(3p_{,X})$ \cite{Tsuji06,AQTW}.
This can be used for dark energy provided that 
$w_{\phi}<-1/3$, i.e. $\lambda^2/p_{,X}<2$.
Unfortunately this condition is incompatible with the constraint 
coming from the BBN. Hence the scaling solution does not exit to
the scalar-field dominated solution in the single-field scenario.

If we consider multiple scalar fields $\phi_i$ ($i=1,2,\cdots, n$)
with the Lagrangian density 
\begin{equation}
p=\sum_{i=1}^n X_i\,g(Y_i)\,,\qquad
Y_i \equiv X_ie^{\lambda_i \phi_i}\,,
\label{psum}
\end{equation}
the scaling solution can be followed by the accelerated
scalar-field dominated point through 
the assisted inflation mechanism.
Even if the individual field does not satisfy the condition 
for inflation, the multiple fields evolve cooperatively to 
give dynamics matching a single-field model 
with \cite{Tsuji06}
\begin{equation}
\frac{1}{\lambda_{\rm eff}^2}=\sum_{i=1}^n
\frac{1}{\lambda_i^2}\,.
\label{lameff}
\end{equation}
Since $\lambda_{\rm eff}$ is reduced compared to 
the individual $\lambda_i$, this allows a possibility to 
exit from the scaling matter era to the regime of 
cosmic acceleration.

In addition to the $n$ scalar fields with the Lagrangian 
density (\ref{psum}) we take into account radiation 
(energy density $\rho_r$) and non-relativistic matter 
(energy density $\rho_m$).
In the flat FLRW space-time with a scale factor $a$
they obey the usual continuity equations 
$\dot{\rho}_r+4H\rho_r=0$ and $\dot{\rho}_m+3H\rho_m=0$, 
respectively, where a dot represents a derivative with respect to 
cosmic time $t$ and $H \equiv \dot{a}/a$ is the Hubble parameter.
The pressure $p_{\phi_i}$ and the energy density $\rho_{\phi_i}$
for the $i$-th scalar field are given, respectively, by 
\begin{eqnarray}
p_{\phi_i} &=& X_i\,g(Y_i)\,,\\
\rho_{\phi_i} &=& 2X_ip_{,X_i}-p_{\phi_i}=
X_i \left[ g(Y_i)+2Y_ig'(Y_i) \right]\,,
\end{eqnarray}
where a prime represents a derivative with respect to $Y_i$.
These satisfy the continuity equation 
\begin{equation}
\dot{\rho}_{\phi_i}+3H(\rho_{\phi_i}+p_{\phi_i})=0\,,
\label{fieldcon}
\end{equation}
which corresponds to 
\begin{eqnarray}
& &\ddot{\phi}_i+3HA(Y_i)p_{,X_i}\dot{\phi}_i  \nonumber \\
& &+\lambda_i X_i \left\{ 1-A(Y_i) [g(Y_i)+
2Y_ig'(Y_i)] \right\}=0\,,
\end{eqnarray}
where 
\begin{equation}
A(Y_i) \equiv \left[ g(Y_i)+5Y_ig'(Y_i)
+2Y_i^2 g'' (Y_i) \right]^{-1}\,.
\end{equation}
The Friedmann equations are 
\begin{eqnarray}
& &3H^2=\sum_{i=1}^n \rho_{\phi_i}+\rho_r+\rho_m\,,
\label{fri1}\\
& &\dot{H}=-\sum_{i=1}^n X_i\,p_{,X_{i}}-\frac23\rho_r
-\frac12 \rho_m
\label{fri2}\,.
\end{eqnarray}

In order to derive autonomous equations we define the 
following quantities
\begin{equation}
x_i \equiv \frac{\dot{\phi}_i}{\sqrt{6}H},\quad
y_i \equiv \frac{e^{-\lambda_i \phi_i/2}}{\sqrt{3}H},\quad
u \equiv \frac{\sqrt{\rho_r}}{\sqrt{3}H}\,,
\end{equation}
where the quantity $Y_i$ defined in Eq.~(\ref{psum}) can be 
expressed as
\begin{equation}
Y_i=x_i^2/y_i^2\,.
\end{equation}
We also introduce the field density parameters
\begin{equation}
\Omega_{\phi_i} \equiv \frac{\rho_{\phi_i}}{3H^2}=
x_i^2 \left[ g(Y_i)+2Y_ig'(Y_i) \right]\,,\quad
\Omega_\phi \equiv \sum_{i=1}^n \Omega_{\phi_i}.
\label{Omephii}
\end{equation}
{}From Eqs.~(\ref{fri1}) and (\ref{fri2}) it follows that 
\begin{eqnarray}
& &\Omega_m \equiv \frac{\rho_m}{3H^2}=
1-\Omega_\phi-\Omega_r\,,
\label{fri1d}\\
& & \frac{\dot{H}}{H^2}=-\frac32-\frac32 \sum_{i=1}^n
x_i^2 g(Y_i)-\frac12 u^2\,,
\label{fri2d}
\end{eqnarray}
where $\Omega_r=u^2$ is the density parameter of radiation.

Using Eqs.~(\ref{fieldcon}) and (\ref{fri2d}), we obtain 
the autonomous equations
\begin{eqnarray}
\hspace*{-2.5em}& &\frac{\rd x_i}{\rd N}= \frac{x_i}{2} \left[ 3+3\sum_{i=1}^n
x_i^2 g(Y_i)+u^2-\sqrt{6}\lambda_i x_i \right] 
\nonumber \\
&&~~~~~~~+\frac{\sqrt{6}}{2}A(Y_i) \left[ \lambda_i \Omega_{\phi_i}
-\sqrt{6} \{ g(Y_i)+Y_ig'(Y_i) \}x_i \right]\,, 
\label{auto1}
\nonumber \\ \\
\hspace*{-2.5em}& & \frac{\rd y_i}{\rd N}= \frac{y_i}{2} \left[ 3+3\sum_{i=1}^n
x_i^2 g(Y_i)+u^2-\sqrt{6}\lambda_ix_i \right]\,,
\label{auto2} \\
\hspace*{-2.5em}& & \frac{\rd u}{\rd N}= \frac{u}{2} \left[-1+3\sum_{i=1}^n
x_i^2 g(Y_i)+u^2 \right]\,,
\label{auto3}
\end{eqnarray}
where $N=\ln\,(a)$.
The field equation of state $w_{\phi_i}$ of the $i$-th field, 
the total field equation of state $w_{\phi}$, 
and the effective equation of state $w_{\rm eff}$
of the system are given, respectively, by
\begin{eqnarray}
& &w_{\phi_i} \equiv \frac{p_{\phi_i}}{\rho_{\phi_i}}=
\frac{g(Y_i)}{g(Y_i)+2Y_ig'(Y_i)}\,, 
\label{wphii} \\
& &w_{\phi} \equiv \frac{\sum_{i=1}^n p_{\phi_i}}
{\sum_{i=1}^n \rho_{\phi_i}}=
\frac{\sum_{i=1}^n x_i^2 g(Y_i)}
{\sum_{i=1}^n x_i^2 [g(Y_i)+2Y_ig'(Y_i)]},
\label{wphidef}
\\
& &w_{\rm eff}\equiv -1-\frac23 \frac{\dot{H}}{H^2}
=\sum_{i=1}^n x_i^2 g(Y_i)+\frac13 u^2.
\end{eqnarray}
%

\section{Fixed points of the system}
\label{fixed}

Let us derive fixed points for the autonomous equations 
(\ref{auto1})-(\ref{auto3}).
In particular we are interested in the scaling solution and 
the scalar-field dominated solution.  
For these solutions the variables $y_i$ do not vanish.
Setting $\rd u/\rd N=0$ in Eq.~(\ref{auto3}), it follows
that $u^2=1-3\sum_{i=1}^n x_i^2 g(Y_i)$ or $u=0$.
The former corresponds to the solution in the presence 
of radiation, whereas the latter to the solution 
without radiation.
In the following we shall discuss these cases separately.

\subsection{Radiation-dominated scaling solution}

Plugging $u^2=1-3\sum_{i=1}^n x_i^2 g(Y_i)$ into 
Eqs.~(\ref{auto1}) and (\ref{auto2}), the fixed point 
for the $i$-th field 
(with $y_i \neq 0$ and $A(Y_i) \neq 0$) satisfies
\begin{equation}
\lambda_i x_i=\frac{2\sqrt{6}}{3}=
\frac{\sqrt{6} \left[g(Y_i)+Y_i g'(Y_i) \right]}
{g(Y_i)+2Y_ig'(Y_i)}\,,
\label{lamxi}
\end{equation}
which gives 
\begin{equation}
Y_ig'(Y_i)=g(Y_i)\,.
\label{gra}
\end{equation}
{}From Eq.~(\ref{wphii}) the field equation of state 
for the $i$-th field is 
\begin{equation}
w_{\phi_i}=1/3\,,
\label{wphii2}
\end{equation}
which means that $\rho_{\phi_i}$ is proportional to $\rho_r$.
Using Eqs.~(\ref{Omephii}) and (\ref{lamxi}) together with 
$p_{,X_i}=g(Y_i)+Y_ig'(Y_i)$, we have 
\begin{equation}
\Omega_{\phi_i}=\frac{4p_{,X_i}}{\lambda_i^2}\,.
\label{Omephii2}
\end{equation}

If all $n$ scalar fields are in the scaling regime, 
$Y_i$ are the same for all $i$ ($Y_i=Y$) from Eq.~(\ref{gra})
and hence $p_{,X_i}$ ($i=1, 2, \cdots, n$) take a
common value $p_{,X}=g(Y)+Yg'(Y)$ with an effective
single-field Lagrangian density $p=Xg(Y)$. 
Then the total field density is given by 
\begin{equation}
\Omega_{\phi}=\frac{4p_{,X}}{\lambda_{\rm eff}^2}\,,
\label{Omephil}
\end{equation}
where $\lambda_{\rm eff}$ is defined in Eq.~(\ref{lameff}).
We are interested in the case where one of the fields, say $\phi_1$,
is in the scaling regime in the deep radiation era, while the energy 
densities of other fields are suppressed relative to that of $\phi_1$.
In the BBN epoch we have the following 
constraint from Eq.~(\ref{Omephii2}):
\begin{equation}
\frac{4p_{,X_1}}{\lambda_1^2} \lesssim 0.045 \quad \to 
\quad \frac{\lambda_1^2}{p_{,X_1}}>88.9\,.
\label{bbn}
\end{equation}

For a given model, i.e. for a given form of $g$, the variables $x_1$ 
and $y_1$ are determined by solving Eq.~(\ref{lamxi}).
If the scalar fields with $i \neq 1$ join the scaling solution
at the late epoch of the radiation era, 
the total field density $\Omega_\phi$ 
tends to increase according to Eq.~(\ref{Omephil}) 
with the decrease of $\lambda_{\rm eff}$.
If the slope $\lambda_2$ of the second scalar field $\phi_2$ 
that joins the scaling solution is of the order of 1, the 
field density (\ref{Omephil}) can be as large as 
$\Omega_\phi=0.1$-1. It is not preferable for many 
fields with low $\lambda_i$ to join the scaling solution
during the radiation era in order to avoid that 
$\Omega_\phi$ exceeds 1.  This can be avoided if the 
field densities $\Omega_{\phi_i}$ ($i \neq 1$) are much smaller 
than the radiation density.

\subsection{Matter-dominated scaling solution and assisted
scalar-field dominated point}

In the absence of radiation ($u=0$) the fixed points for the $i$-th field 
corresponding to $y_i \neq 0$ and $A(Y_i) \neq 0$ obey 
the following equations
\begin{eqnarray}
& & 3+3\sum_{i=1}^n x_i^2g(Y_i)=\sqrt{6}\lambda_i x_i\,,
\label{meq1} \\
& & \lambda_i \Omega_{\phi_i}=\sqrt{6} [g(Y_i)+Y_ig'(Y_i)]x_i\,.
\label{meq2}
\end{eqnarray}
{}From Eqs.~(\ref{Omephii}), (\ref{wphii}), (\ref{meq1}) 
and (\ref{meq2}) it follows that 
\begin{equation}
w_{\phi_i}=\sum_{i=1}^n x_i^2 g(Y_i)
=-1+\frac{\sqrt{6}}{3}\lambda_i x_i\,.
\label{wphiie}
\end{equation}
Since $w_{\phi_i}\Omega_{\phi_i}=x_i^2 g(Y_i)$ we have 
\begin{equation}
w_{\phi_i}=\sum_{i=1}^n w_{\phi_i} \Omega_{\phi_i}\,.
\label{mwphire}
\end{equation}
In the case of a single field $\phi_i$, this equation gives
$w_{\phi_i}=0$ or $\Omega_{\phi_i}=1$.
The former corresponds to the scaling solution along 
which $\rho_{\phi_i}$ is proportional to the matter 
density $\rho_m$, whereas the latter is 
the scalar-field dominated solution.

If all $n$ scalar fields are on the fixed points characterized by 
the condition (\ref{wphiie}), 
it follows that $w_{\phi_1}=\cdots=w_{\phi_i}=\cdots=
w_{\phi_n} \equiv w_{\phi}$
and hence $Y_1=\cdots=Y_i=\cdots =Y_n \equiv Y$ from 
Eq.~(\ref{wphii}). In this case one has either $w_{\phi}=0$
or $\Omega_{\phi}=1$ from Eq.~(\ref{mwphire}).
Equation (\ref{wphiie}), which holds for the each scalar field,
reduces to the single-field system
\begin{equation}
w_{\phi}=x^2 g(Y)
=-1+\frac{\sqrt{6}}{3}\lambda_{\rm eff} x \,,
\label{wphis}
\end{equation}
where $x=\lambda_i x_i/\lambda_{\rm eff}$.
The effective single-field Lagrangian density is given by $p=Xg(Y)$
with $\Omega_\phi=x^2 \left[ g(Y)+2Yg'(Y) \right]$.
We also note that Eqs.~(\ref{meq1}) and (\ref{meq2}) 
reduce to the following effective single-field forms:
\begin{eqnarray}
& & 3+3x^2g(Y)=\sqrt{6}\lambda_{\rm eff} x\,,
\label{meq1s} \\
& & \lambda_{\rm eff}\Omega_{\phi}=\sqrt{6}p_{,X}x\,,
\label{meq2s}
\end{eqnarray}
where $p_{,X}=g(Y)+Yg'(Y)$.

In the following we shall discuss the matter-dominated 
scaling solution and the assisted field-dominated 
solution, separately.

\subsubsection{Matter-dominated scaling solution}

If the $i$-th scalar field is in the scaling regime during 
the matter-dominated epoch, i.e. $w_{\phi_i}=0$,
it follows from Eqs.~(\ref{wphii}) 
and (\ref{wphiie}) that $\lambda_i x_i=\sqrt{6}/2$ and
\begin{equation}
g(Y_i)=0\,.
\label{gm}
\end{equation}
{}From Eq.~(\ref{Omephii}) we obtain 
\begin{equation}
\Omega_{\phi_i}=\frac{3p_{,X_i}}{\lambda_i^2}\,.
\label{mOmep}
\end{equation}
More generally the field density parameter in the presence of a
perfect fluid with an equation of state $w_{f}$ is given by 
$\Omega_{\phi_i}=3(1+w_{f})p_{,X_i}/\lambda_i^2$ \cite{Tsuji06}.

If all $n$ scalar fields are in the scaling regime, then they can be 
described by an effective single-field system with $w_{\phi}=0$ and 
\begin{equation}
\Omega_{\phi}=\frac{3p_{,X}}{\lambda_{\rm eff}^2}\,.
\end{equation}
This scaling solution is stable for 
$\lambda_{\rm eff}^2>3p_{,X}$ \cite{Tsuji06}.

\subsubsection{Assisted field-dominated point}

Besides the matter scaling solution discussed above, there is another
fixed point that can be responsible for the late-time acceleration.
In the single-field case the solutions do not exit to the accelerated 
field-dominated point ($\Omega_{\phi_i}=1$) 
from the scaling matter era, because the scaling solution is stable 
for $\Omega_{\phi_i}=3p_{,X_i}/\lambda_i^2<1$.
However the presence of multiple scalar fields allows this transition.

Since $\Omega_\phi=1$ in Eq.~(\ref{meq2s}) for the
scalar-field dominated point with $n$ multiple fields, 
it follows from Eq.~(\ref{wphis}) that 
\begin{equation}
w_{\phi}=-1+\frac{\lambda_{\rm eff}^2}{3p_{,X}}\,.
\label{assiat}
\end{equation}
This fixed point can be responsible for the late-time 
acceleration ($w_{\phi}<-1/3$) for $\lambda_{\rm eff}^2<2p_{,X}$.
Moreover it is stable under the condition 
$\lambda_{\rm eff}^2<3p_{,X}$ \cite{Tsuji06} 
(which is opposite to the stability of the scaling matter solution).
Using the relations $w_{\phi}\Omega_{\phi_i}=x_i^2 g(Y)$ 
and $w_{\phi}=x^2g(Y)$, we find 
\begin{equation}
\Omega_{\phi_i}=\frac{x_i^2}{x^2}=\frac{\lambda_{\rm eff}^2}
{\lambda_i^2}\,.
\label{Omephire}
\end{equation}

We shall study the case in which one of the fields has a large
slope $\lambda_1~(\gg 1)$ to satisfy the BBN bound (\ref{bbn})
and other fields with $\lambda_i={\cal O}(1)$
join the scalar-field dominated attractor
(\ref{assiat}) at late times. 
Since the joining of such multiple scalar fields reduces
$\lambda_{\rm eff}$ it should be possible to give rise
to sufficient cosmic acceleration through the assisted 
inflation mechanism, even if the individual field cannot 
be responsible for the acceleration.

\vspace{0.5cm}

For a given model one can derive $Y_1$ (for the 
field $\phi_1$) that corresponds to the scaling solution
during radiation and matter eras by solving 
Eqs.~(\ref{gra}) and (\ref{gm}), respectively.
The field density parameters $\Omega_{\phi_1}$ in these epochs
are given by Eqs.~(\ref{Omephii2}) and (\ref{mOmep}), respectively.
The assisted field-dominated solution corresponds to 
\begin{equation}
\frac{6\left[ g(Y)+Yg'(Y) \right]^2}
{g(Y)+2Yg'(Y)}=\lambda_{\rm eff}^2\,,
\label{gYre}
\end{equation}
which comes from by combining Eqs.~(\ref{meq1s}) and (\ref{meq2s})
with $\Omega_{\phi}=1$. By solving this equation for a given form of $g(Y)$, 
we obtain the field equation of state (\ref{assiat}) and also 
$x=\lambda_{\rm eff}/(\sqrt{6}p_{,X})$ from Eq.~(\ref{meq2s}).

In subsequent sections we shall consider two models: 
(i) quintessence with multiple exponential potentials, and 
(ii) the multi-field dilatonic ghost condensate model 
(one of k-essence models).
In our numerical simulations we identify the present epoch 
(the redshift $z=0$) to be $\Omega_{\phi}=0.72$
with the radiation density in the region 
$7.0 \times 10^{-5}<\Omega_r<1.0 \times 10^{-4}$.

\section{Quintessence with multiple exponential potentials}
\label{exponential}

The single-field quintessence with an exponential potential corresponds to 
the Lagrangian density $p=X-ce^{-\lambda \phi}$, i.e.
the choice $g(Y)=1-c/Y$ in Eq.~(\ref{plagform}).
In the following we shall consider the Lagrangian density (\ref{psum})
of $n$ scalar fields with the choice $g(Y_i)=1-c_i/Y_i$
($i=1,2,\cdots, n$).

Since $p_{,X_i}=g(Y_i)+Y_ig'(Y_i)=1$ in this model the scaling 
field density $\Omega_{\phi_i}$ during the radiation and matter eras is given by 
$\Omega_{\phi_i}=4/\lambda_i^2$ and $\Omega_{\phi_i}=3/\lambda_i^2$,  
respectively [see Eqs.~(\ref{Omephii2}) and (\ref{mOmep})].
Below we discuss the case in which one of the scalar fields, $\phi_1$, is 
in the scaling regime during most of the radiation and matter eras and 
other fields eventually join the assisted scalar-field dominated 
attractor with $w_{\phi}$ given by Eq.~(\ref{assiat}). 
Then the BBN bound (\ref{bbn}) gives 
\begin{equation}
\lambda_1>9.42\,.
\label{bbnexp}
\end{equation}

Under this condition, the scaling field density $\Omega_{\phi_1}$
during the matter-dominated epoch is constrained to be 
$\Omega_{\phi_1}<0.034$.
If other fields join the scaling regime in the radiation 
(matter) era, the field density increases from $\Omega_{\phi_1}=4/\lambda_1^2$
($\Omega_{\phi_1}=3/\lambda_1^2$) to $\Omega_{\phi}=4/\lambda_{\rm eff}^2$
($\Omega_{\phi}=3/\lambda_{\rm eff}^2$).
This is possible provided that the slopes of the joining scalar fields
satisfy the conditions $\lambda_i \gg 1$. 
Meanwhile, if $\lambda_i$ are of the order of 1, this leads to a large 
density parameter $\Omega_\phi$ that is comparable to unity. 
In what follows we focus on the case in which the fields with slopes
$\lambda_i={\cal O}(1)$ ($i \ge 2$) enter the regime of 
the assisted cosmic acceleration preceded by 
scaling solutions induced by $\phi_1$. 

It is convenient to introduce the following variable
\begin{equation}
\tilde{y}_i \equiv \sqrt{c_i}\,y_i\,.
\end{equation}
{}From Eq.~(\ref{lamxi}) the radiation-dominated scaling solution 
for the field $\phi_1$ corresponds to 
\begin{equation}
(x_1, \tilde{y}_1)=\left( \frac{2\sqrt{6}}{3\lambda_1}, 
\frac{2\sqrt{3}}{3\lambda_1} \right)\,,\qquad
Y_1=2c_1\,.
\label{rafix}
\end{equation}
This is followed by the matter-dominated scaling 
solution, satisfying
\begin{equation}
(x_1, \tilde{y}_1)=\left( \frac{\sqrt{6}}{2 \lambda_1}, 
\frac{\sqrt{6}}{2\lambda_1} \right)\,,\qquad
Y_1=c_1\,.
\label{mafix}
\end{equation}
The assisted field-dominated point corresponds to the single-field 
potential $V(\phi)=ce^{-\lambda_{\rm eff}\phi}$, 
i.e. $g(Y)=1-c/Y$ with $Y=x^2/y^2$.
{}From Eqs.~(\ref{gYre}) and (\ref{meq2s}) this is characterized 
by the fixed point (where we define $\tilde{y} \equiv \sqrt{c}y$):
\begin{equation}
(x, \tilde{y})=\left( \frac{\lambda_{\rm eff}}{\sqrt{6}}, 
\sqrt{1-\frac{\lambda_{\rm eff}^2}{6}} \right)\,,\qquad
Y=\frac{\lambda_{\rm eff}^2}{6-\lambda_{\rm eff}^2}c\,.
\label{assifix}
\end{equation}
For the  $i$-th field we have that $x_i=(\lambda_{\rm eff}/\lambda_i)x$
and $y_i=x_i/\sqrt{Y}$.

\subsection{Two fields}

First let us consider the case of two scalar fields
$\phi_1$ and $\phi_2$.

\begin{figure}
\includegraphics[height=3.5in,width=3.5in]{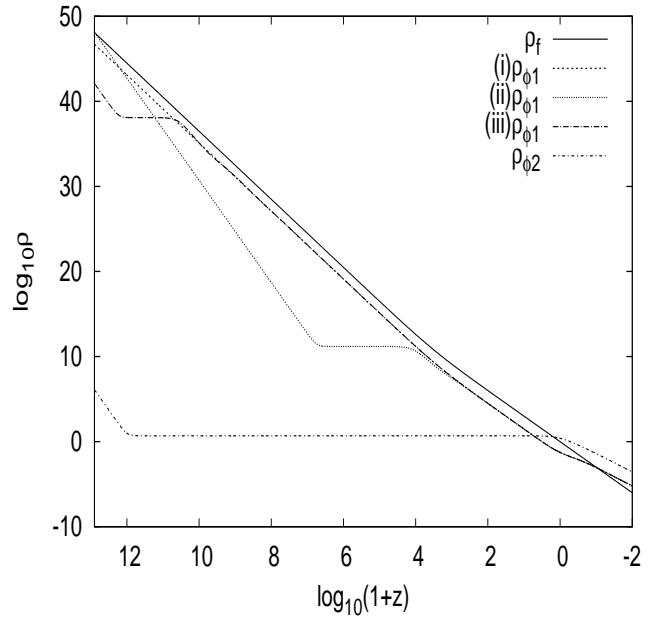}
\caption{\label{fig1}
Evolution of the energy densities $\rho_f=\rho_r+\rho_m$, 
$\rho_{\phi_1}$, and $\rho_{\phi_2}$ versus the redshift $z$
for $\lambda_1=10$ and $\lambda_2=1.5$
in the two-field quintessence with exponential potentials.
We choose three different initial conditions for $\phi_1$: 
(i) $x_1=2\sqrt{6}/(3\lambda_{1})$, 
$\tilde{y}_1=2\sqrt{3}/(3\lambda_{1})$, 
(ii) $x_1=0.99$, $\tilde{y}_1=0.12$, and 
(iii) $x_1=1\times 10^{-3}$, $\tilde{y}_1=1\times10^{-5}$,
while other initial conditions are fixed to be
$x_2=1\times 10^{-21}$, $\tilde{y}_2=2\times 10^{-24}$, 
and $\Omega_m=4\times 10^{-10}$.
In the case (i) the field $\phi_1$ is in the scaling 
regime from the beginning.}
\end{figure}

In Fig.~\ref{fig1} we plot the evolution of the background 
fluid density $\rho_f=\rho_r+\rho_m$ and the field densities
$\rho_{\phi_1}, \rho_{\phi_2}$ versus the redshift 
$z=a_0/a-1$ ($a_0$ is the present value of $a$) 
for $\lambda_1=10$ and $\lambda_2=1.5$.
We choose three different initial conditions for $\phi_1$.
The case (i) corresponds to the exact scaling solution 
starting from the fixed point (\ref{rafix}),
along which $\Omega_{\phi_1}=4/\lambda_1^2=0.04$
and $\Omega_{\phi_1}=3/\lambda_1^2=0.03$ during 
radiation and matter eras, respectively.
Finally the system enters the epoch in which the energy
density $\rho_{\phi_2}$ of the second field $\phi_2$
dominates the dynamics.
Figure \ref{fig1} shows that the field $\phi_1$ eventually joins
the scaling regime both for the initial conditions
(ii) $\rho_{\phi_1} \approx \rho_f$ and 
(iii) $\rho_{\phi_1} \ll \rho_f$.
Thus the cosmological trajectories converge to a common 
scaling solution for a wide range of initial conditions.

\begin{figure}
\includegraphics[height=3.4in,width=3.5in]{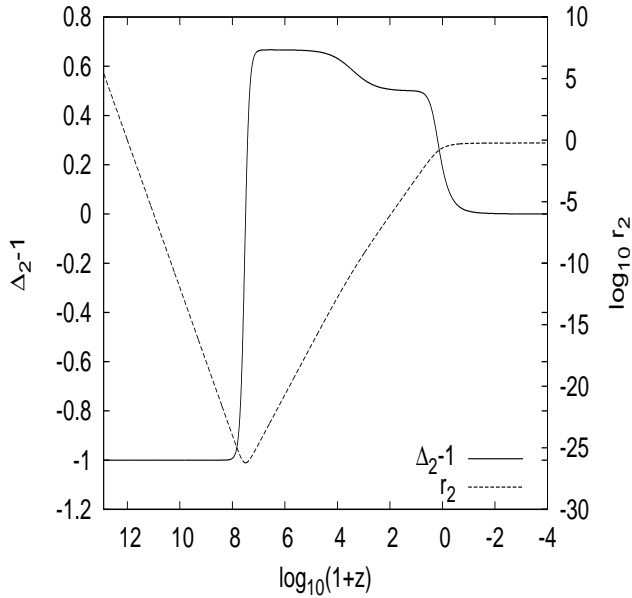}
\caption{\label{fig2}
Evolution of $r_2$ and $\Delta_2-1$ 
for $\lambda_1=10$ and $\lambda_2=1.5$ 
in the two-field quintessence with exponential potentials.
The same initial conditions are chosen as in the case (i) 
of Fig.~\ref{fig1}. Initially the quantity $r_2$
decreases as $r_2 \propto e^{-6N}$ because $\Delta_2 \approx 0$.
The ratio $r_2$ starts to increase after $\Delta_2$ 
becomes larger than 1.}
\end{figure}
\begin{figure}
\includegraphics[height=3.4in,width=3.5in]{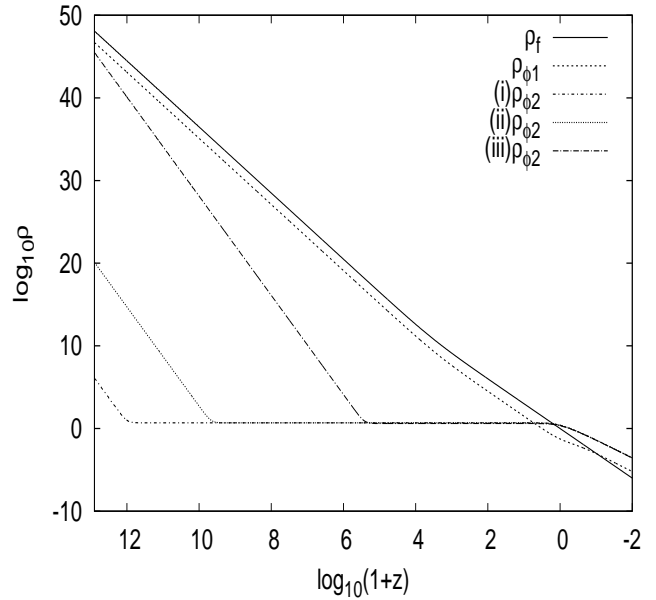}
\caption{\label{fig3}
Evolution of $\rho_f$, $\rho_{\phi_1}$, and $\rho_{\phi_2}$ 
for $\lambda_1=10$ and $\lambda_2=1.5$
in the two-field quintessence with exponential potentials
with three different initial conditions 
for the kinetic energy of $\phi_2$: 
(i) $x_2=1\times 10^{-21}$, (ii) $x_2=1\times 10^{-14}$, 
and (iii) $x_2=0.05$. Other initial conditions 
are chosen to be $\tilde{y}_2=2\times 10^{-24}$,
$x_1=2\sqrt{6}/(3\lambda_{1})$, 
$\tilde{y}_1=2\sqrt{3}/(3\lambda_{1})$, and 
$\Omega_m=4\times 10^{-10}$.
The field $\phi_2$ enters the regime with a nearly 
constant $\rho_{\phi_2}$ independent of its initial kinetic energy. }
\end{figure}
\begin{figure}
\includegraphics[height=3.5in,width=3.5in]{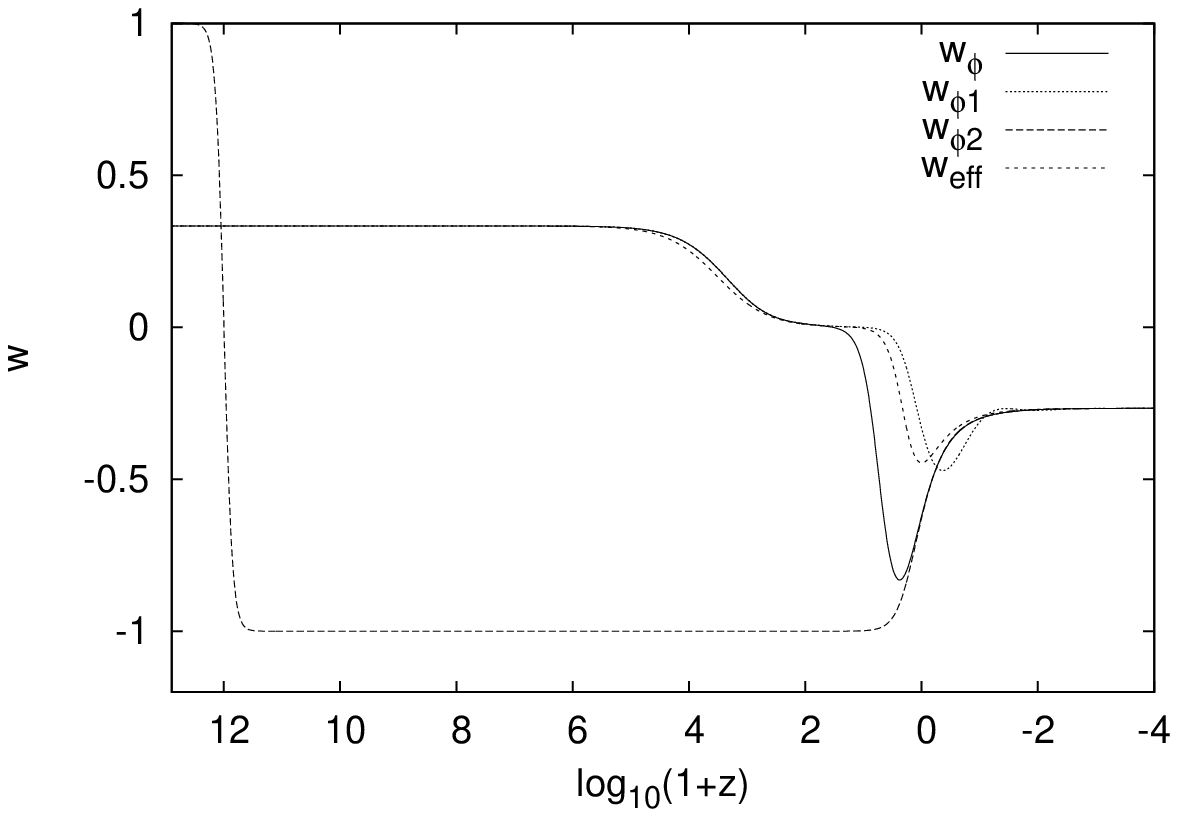}
\caption{\label{fig4}
Evolution of the field equations of state $w_{\phi}$, 
$w_{\phi_1}$, $w_{\phi_2}$, and the effective 
equation of state $w_{\rm eff}$ for 
$\lambda_1=10$ and $\lambda_2=1.5$ 
in the two-field quintessence with exponential potentials.
The initial conditions are chosen to be
$x_1=2\sqrt{6}/(3\lambda_{1})$, 
$\tilde{y}_1=2\sqrt{3}/(3\lambda_{1})$, 
$x_2=1\times 10^{-21}$, 
$\tilde{y}_2=2\times 10^{-24}$, and 
$\Omega_m=4\times 10^{-10}$.}
\end{figure}

In Fig.~\ref{fig1} we find that the second field density 
$\rho_{\phi_2}$ is almost frozen after the initial transient 
period. In order to understand this behavior we introduce 
the ratio $r_i$ between the kinetic energy $\dot{\phi}_i^2/2$
and the potential energy $V_i (\phi_i)=c_i e^{-\lambda_i \phi_i}$
of the $i$-th field:
\begin{equation}
r_i \equiv \frac{\dot{\phi}_i^2}{2V_i}\,,
\end{equation}
which is related to the quantity $Y_i$ via $r_i=Y_i/c_i$.
Taking the derivative of $r_i$ with respect to $N$, 
it follows that \cite{tracking}
\begin{equation}
\frac{\rd \ln r_i}{\rd N}=6 \left[ \Delta_i (t) -1 \right],\qquad
\Delta_i (t) \equiv \lambda_i \sqrt{\frac{\Omega_{\phi_i}}
{3(1+w_{\phi_i})}}.
\label{Del}
\end{equation}

For the scaling field $\phi_1$ one has $w_{\phi_1}=w_f$ and 
$\Omega_{\phi_1}=3(1+w_f)/\lambda_1^2$ \cite{CLW}, 
where $w_f$ is the equation of state of the background fluid.
This means that $\Delta_1(t)=1$, so that the ratio $r_1=Y_1/c_1$
remains constant. In fact, from Eqs.~(\ref{rafix}) and (\ref{mafix}), 
one has $r_1=2$ and $r_1=1$ during the radiation and matter eras, 
respectively. This reflects the fact that the scaling field 
has a kinetic energy with the same order as its potential energy. 

The field $\phi_2$ joining the assisted attractor at late-times 
satisfies $\lambda_2={\cal O}(1) \ll \lambda_1$ and 
$\Omega_{\phi_2} \ll \Omega_{\phi_1}$
at the early stage of the radiation era, so that $\Delta_2 (t) \ll 1$
initially (unless $w_{\phi_i}$ is unnaturally close to $-1$).
At this stage the ratio $r_2$ decreases rapidly as $\propto e^{-6N}$
according to Eq.~(\ref{Del}), see Fig.~\ref{fig2}.
In the region $r_2 \ll 1$ the field $\phi_2$ is almost frozen 
with nearly constant $\rho_{\phi_2}$.
As $r_2$ decreases, $\Omega_{\phi_2}$ grows and $w_{\phi_2}$
approaches $-1$. This leads to the growth of $\Delta_2(t)$.
As we see in Fig.~\ref{fig2} the ratio $r_2$ starts to 
increase after $\Delta_2 (t)$ becomes larger than 1.
When $r_2$ grows to the order of 1, the field $\phi_2$
begins to evolve to join the assisted 
attractor given by Eq.~(\ref{assifix}).

The mass squared for the $i$-th scalar field is given by 
$m_i^2 \equiv \rd^2 V_i (\phi_i)/\rd \phi_i^2=
\lambda_i^2 V_i (\phi_i)$.
The energy density of $\phi_2$ starts to dominate 
around the present epoch, so that 
$3H_0^2 \approx V_2 (\phi_2^{(0)})$
(the subscript ``0'' represents present values).
Then the mass of $\phi_2$ can be estimated as
\begin{equation}
m_2 (\phi_2^{(0)}) \approx \lambda_2 H_0\,.
\label{mass}
\end{equation}
Recall that $\lambda_2$ needs to be of the order of 1 
to realize a stable assisted attractor satisfying the condition 
$\lambda_{\rm eff}^2<3$.
Hence the mass $m_2$ is as small as $H_0$ today.
In the numerical simulations of Figs.~\ref{fig1} and \ref{fig2}
the field $\phi_2$ is almost frozen with the mass (\ref{mass})
for the redshift $1 \lesssim z \lesssim 10^{12}$
(during which the condition $r_2 \ll 1$ is fulfilled).

Even if the field $\phi_2$ is rapidly rolling at the initial stage of 
the radiation era such that $r_2 \gg 1$, it enters 
the regime in which $\phi_2$ is nearly frozen 
($w_{\phi_2} \simeq -1$) prior to the matter-dominated epoch.
In Fig.~\ref{fig3} the evolution of $\rho_{\phi_2}$ is plotted 
for three different initial conditions of $x_2$ 
with $\tilde{y}_2$ fixed.
The dominance of the field kinetic energy relative to its potential energy 
corresponds to $r_2 \gg 1$ and $w_{\phi_2} \simeq 1$, 
which results in the rapid decrease of $\Omega_{\phi_2}$
to reach the regime $\Delta_2(t) \ll 1$.
Even if $r_2 \gg 1$ initially, the decrease of $r_2$ in 
the regime $\Delta_2(t) \ll 1$ is so fast ($\propto e^{-6N}$) that 
the field $\phi_2$ eventually enters the frozen regime
with $w_{\phi_2} \simeq -1$.
For the initial conditions satisfying $r_2 \ll 1$ the field $\phi_2$ is 
almost frozen from the beginning, so that 
$\rho_{\phi_2}$ is nearly constant until recently.

If we change the initial conditions of $\tilde{y}_2$ 
associated with the field potential, this leads to the modification 
of the epoch at which the field $\phi_2$ dominates at late times.
This comes from the fact that the density $\rho_{\phi_2}$
during which $\phi_2$ is nearly frozen is sensitive to the 
choice of its initial potential energy. 
Thus the evolution of the field $\phi_2$ depends on its
initial potential energy but not on its initial kinetic energy.

Figure \ref{fig4} illustrates the variation of $w_{\phi}$, 
$w_{\phi_1}$, $w_{\phi_2}$, and $w_{\rm eff}$
for $\lambda_1=10$ and $\lambda_2=1.5$
with the same initial condition as in the case (i) of Fig.~\ref{fig1}.
The equations of state $w_{\phi}$ and $w_{\phi_1}$ 
are similar to the effective equation of state $w_{\rm eff}$
during radiation and matter eras, but the deviation appears
at low redshifts. The field $\phi_2$ is almost frozen 
around $w_{\phi_2}=-1$ after the initial transient period, 
but it begins to evolve for $z \lesssim {\cal O}(1)$.

{}From the definition of $w_{\phi}$ in 
Eq.~(\ref{wphidef}) we have
\begin{equation}
w_{\phi}=\frac{1}{\Omega_{\phi}}
\left( w_{\phi_1} \Omega_{\phi_1}+w_{\phi_2} 
\Omega_{\phi_2} \right)\,.
\end{equation}
Note that $w_{\phi_1} \approx 0$ and $w_{\phi_2} \approx -1$ 
around the end of the matter-dominated epoch. 
After $\Omega_{\phi_2}$ gets 
larger than $\Omega_{\phi_1}$, $w_{\phi}$ begins to be 
mainly determined by the field $\phi_2$, i.e. 
$w_{\phi} \approx w_{\phi_2} \Omega_{\phi_2}/\Omega_{\phi}$.
As we see in Fig.~\ref{fig4} $w_{\phi}$ takes a minimum 
before reaching the present epoch ($z=0$), which
is followed by its increase toward the 
attractor value $w_{\phi}=-1+\lambda_{\rm eff}^2/3$.
For the model parameters used in the numerical simulation of 
Fig.~\ref{fig4} we have $\lambda_{\rm eff}=1.483$, 
which gives $w_{\phi}=-0.267$ at the scalar-field dominated attractor.
This corresponds to the decelerated expansion of the universe.
Meanwhile one has $w_{\phi}(z=0)=-0.62$ and $w_{\rm eff}(z=0)=-0.45$, 
which means that the transient acceleration 
occurs at the present epoch.
Interestingly, even without the assisted accelerated attractor, 
such a temporal acceleration can be realized by the 
presence of the thawing field $\phi_2$.

\begin{figure}
\includegraphics[height=3.2in,width=3.5in]{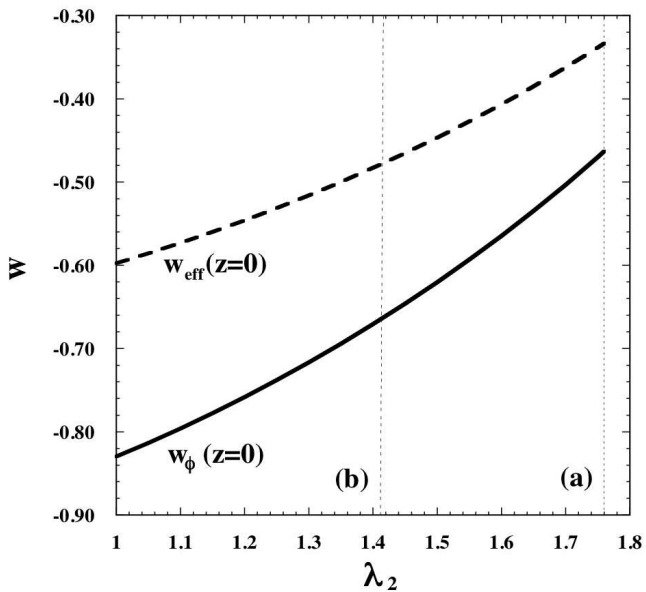}
\caption{\label{fig5}
The field equation of state $w_{\phi}$ today versus $\lambda_2$
for $\lambda_1=9.43$ (solid curve)
in the two-field quintessence with exponential potentials.
If $\lambda_2<1.76$ the scalar-field dominated 
point is the final attractor. 
The condition for cosmic acceleration ($w_{\rm eff} (z=0)<-1/3$) 
is satisfied even for $\lambda_2>\sqrt{2}$.
The two lines (a) and (b) in the figure correspond to $\lambda_2=1.76$
and $\lambda_2=\sqrt{2}$, respectively.}
\end{figure}

Under the BBN bound (\ref{bbnexp}) and the condition 
$\lambda_2>\sqrt{2}$ (i.e. the field $\phi_2$ cannot be responsible for 
the accelerated expansion as a single component of the universe),
the equation of state $w_{\phi}$ for the late-time assisted attractor is
not very different from $-1/3$. Meanwhile the present value of $w_{\phi}$ 
is smaller than its asymptotic value.
For the marginal case with $\lambda_1=9.43$ and $\lambda_2=1.415$
we find that $w_{\phi}(z=0)=-0.66$ numerically. 
For increasing $\lambda_2$ we obtain larger values of $w_{\phi}(z=0)$
and $w_{\rm eff}(z=0)$, as we see in Fig.~\ref{fig5}.
If we do not impose the condition $\lambda_2>\sqrt{2}$,
then $w_{\phi}(z=0)$ can be smaller than $-0.66$.
Note that, when $\lambda_2>1.76$ and $\lambda_1=9.43$, 
the scalar-field dominated point ceases to be the late-time attractor.
We have also carried out numerical simulations for different values of 
$\lambda_1$ satisfying the condition $\lambda_1>9.42$ and found 
that $w_{\phi}(z=0)$ and $w_{\rm eff}(z=0)$ are insensitive to 
the change of $\lambda_1$.

\subsection{More than two fields}

For three scalar fields the total field equation of state is given by  $w_{\phi}=(w_{\phi_1}\Omega_{\phi_1}+w_{\phi_2}\Omega_{\phi_2}+
w_{\phi_3}\Omega_{\phi_3})/\Omega_{\phi}$.
If the two fields $\phi_2$ and $\phi_3$ with slopes 
$\lambda_2, \lambda_3={\cal O}(1)$
join the assisted attractor for $z \lesssim {\cal O}(1)$, 
it is possible to obtain smaller values of $w_{\phi}(z=0)$ and 
$w_{\rm eff}(z=0)$ relative to the two-field case.

\begin{figure}
\includegraphics[height=3.5in,width=3.5in]{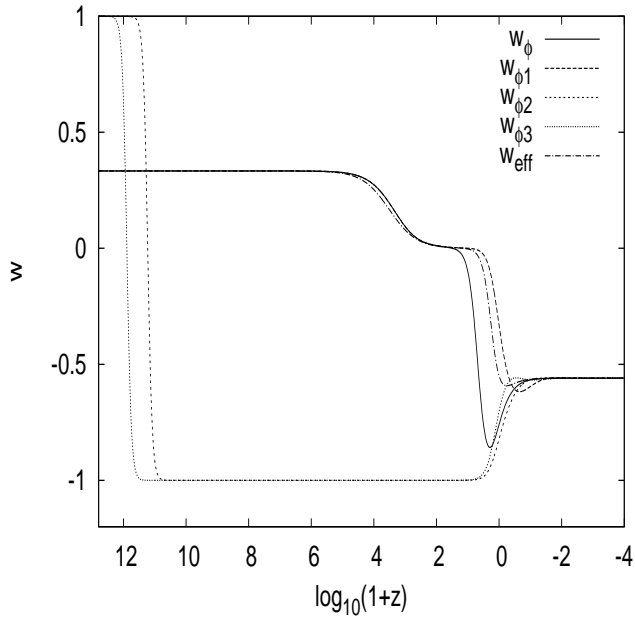}
\caption{\label{fig6}
Evolution of $w_{\phi}$, $w_{\phi_1}$, $w_{\phi_2}$, 
$w_{\phi_3}$, and $w_{\rm eff}$ for $\lambda_1=9.5$, 
$\lambda_2=1.42$, and $\lambda_3=2.0$ 
in the three-field quintessence with exponential potentials.
The initial conditions are chosen to be
$x_1=2\sqrt{6}/(3\lambda_{1})$, 
$\tilde{y}_1=2\sqrt{3}/(3\lambda_{1})$, 
$x_2=1\times 10^{-19}$, 
$\tilde{y}_2=2\times 10^{-24}$, 
$x_3=1\times 10^{-21}$, 
$\tilde{y}_3=2\times 10^{-24}$, and 
$\Omega_m=5\times 10^{-10}$.}
\end{figure}

In Fig.~\ref{fig6} we plot the evolution of $w_{\phi}$, $w_{\phi_i}$ ($i=1, 2, 3$) 
as well as $w_{\rm eff}$ for $\lambda_1=9.5$, $\lambda_2=1.42$, and 
$\lambda_3=2.0$.
The field $\phi_1$ is in the scaling regime during the radiation and matter eras, 
which is followed by the epoch of cosmic acceleration once the energy 
densities of $\phi_2$ and $\phi_3$ are dominant.
The fields $\phi_2$ and $\phi_3$ have been nearly frozen 
(except for the initial transient period) 
by the time they start to evolve for $z \lesssim {\cal O}(1)$. 
In the numerical simulation of Fig.~\ref{fig6} the energy densities $\rho_{\phi_2}$
and $\rho_{\phi_3}$ are the same order 
when they begin to dominate over the background fluid density. 
In the numerical simulation of Fig.~\ref{fig6} the field equation of state today is
found to be $w_{\phi}(z=0)=-0.76$, which is smaller than the minimum value 
$w_{\phi}(z=0)=-0.66$ in the two-field case.
This comes from the fact that the third field with $w_{\phi_3}$ close to
$-1$ leads to smaller values of  $w_{\phi}(z=0)$.

For the marginal model parameters $\lambda_1=9.43$, $\lambda_2=\lambda_3=1.415$, 
which satisfy the conditions $\lambda_1>9.42$ and $\lambda_2, \lambda_3>\sqrt{2}$,
we find that $w_{\phi}(z=0)=-0.83$, provided the fields $\phi_2$ and $\phi_3$ 
exit from the frozen regime almost at the same time. 
If either $\phi_2$ or $\phi_3$ begins to evolve much later than another,
then $w_{\phi}(z=0)$ tends to be larger. 
This case is not much different from the two-field scenario for estimating 
the value of $w_{\phi}(z=0)$, although the scalar-field dominated attractor 
is different. In the three-field scenario the equation of state $w_{\phi}$ 
for the attractor can be as small as $w_{\phi} \sim -0.6$
for $\lambda_2, \lambda_3={\cal O}(1)$ so that 
cosmic acceleration today is not transient.

\begin{figure}
\includegraphics[height=3.5in,width=3.5in]{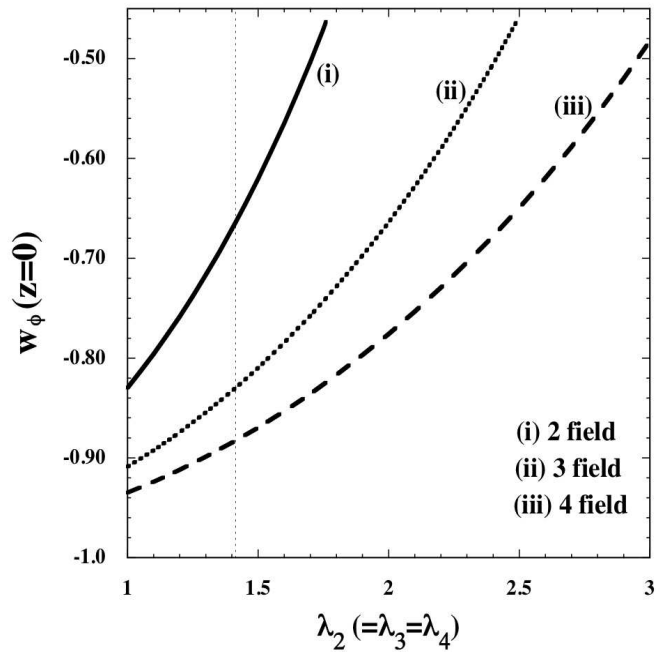}
\caption{\label{fig7}
The equation of state $w_{\phi}$ today 
versus $\lambda_i$ ($i \neq 1$) for $\lambda_1=9.43$
in the multi-field quintessence with exponential potentials.
The slopes $\lambda_i$ for the $i$-th ($i \ge 2$) 
fields are chosen to be the same.
In this simulation the fields $\phi_i$ ($i \ge 2$) enter 
the regime of the assisted field-dominated attractor 
almost at the same time.
Under the condition $\lambda_i>\sqrt{2}$ (which
is shown as a thin dotted line in the figure), we have 
that $w_{\phi}<-0.8$ for more than two fields. }
\end{figure}

In Fig.~\ref{fig7} we show $w_{\phi}(z=0)$ versus $\lambda_i$ ($i \ge 2$)
for $\lambda_1=9.43$ in the presence of multiple scalar fields.
Under the condition $\lambda_i >\sqrt{2}$ the observational 
bound, $w_{\phi}(z=0)<-0.8$, can be satisfied for three fields.
In the case of four fields it is possible to satisfy the same 
bound for $\lambda_i<2$ ($i=2,3,4$). As we add more fields, we obtain 
smaller values of $w_{\phi}(z=0)$ and $w_{\rm eff}(z=0)$.

\section{Multi-field dilatonic ghost condensate model}
\label{dilatonic}

Let us next proceed to the dilatonic ghost 
condensate model with $n$ scalar fields, 
where the Lagrangian density is given by 
(\ref{psum}) with the choice $g(Y_i)=-1+c_iY_i$ ($i=1, 2, \cdots, n$), 
i.e. $p=\sum_{i=1}^n (-X_i+c_i e^{\lambda_i \phi_i}X_i^2)$.
The coefficients $c_i$ are positive so that the quantum 
instability problem of the negative kinetic energy $(-X_i)$
can be avoided by the presence of the higher-order
derivative term $c_i e^{\lambda_i \phi_i}X_i^2$ \cite{PT04}.

In this model we have $p_{,X_i}=2\tilde{Y}_i-1$ and 
\begin{equation}
w_{\phi_i}=\frac{\tilde{Y}_i-1}{3\tilde{Y}_i-1}\,,\quad
\Omega_{\phi_i}=x_i^2 (3\tilde{Y}_i-1)\,,
\label{wphigho}
\end{equation}
where 
\begin{equation}
\tilde{Y}_i \equiv c_i Y_i\,.
\end{equation}
When $\tilde{Y}_i=1/2$ the equation of state $w_{\phi_i}$
is equivalent to $-1$. 
The quantum stability of the scalar field is ensured
for $\tilde{Y}_i \ge 1/2$ (i.e. $w_{\phi_i} \ge -1$), 
whereas in the region $\tilde{Y}_i < 1/2$ the vacuum 
is unstable against the catastrophic particle production
of ghost and normal fields \cite{PT04}. 
In the following we shall focus on the case $\tilde{Y}_i \ge 1/2$.

{}From Eqs.~(\ref{auto1}) and (\ref{auto2}) we obtain 
the following equations for $\tilde{Y}_i$:
\begin{equation}
\frac{\rd \tilde{Y}_i}{\rd N}=\tilde{Y}_i 
\frac{3\tilde{Y}_i (\sqrt{6}\lambda_i x_i-4)+
6-\sqrt{6}\lambda_i x_i}{6\tilde{Y}_i-1}\,,
\label{auto4}
\end{equation}
which hold for $i=1, 2, \cdots, n$.
We will solve Eqs.~(\ref{auto1}), (\ref{auto3}) and (\ref{auto4})
in our numerical simulations.

For this model the solution to Eq.~(\ref{gra}) does not exist, 
whereas the solution to Eq.~(\ref{gm}) is given by 
$\tilde{Y}_i=1$. This means that the scaling solution
is absent during the radiation era, while it is present 
during the matter era. More precisely, for the background fluid
with an equation of state $w_f$, the presence of the 
scaling solution corresponds to the condition 
$(1-w_f)g(Y_i)=2w_f Y_ig'(Y_i)$ \cite{Tsuji06}.
Solving this equation for the present model, 
we obtain
\begin{equation}
\tilde{Y}_i=\frac{1-w_f}{1-3w_f}\,.
\end{equation}

For the radiation fluid ($w_f=1/3$) we require that $\tilde{Y}_i \to \infty$
for the existence of the scaling solution.
If the field $\phi_1$ is in a nearly scaling regime during 
the radiation era, it follows that $\Omega_{\phi_1} \simeq 
(4/\lambda_1^2)(2\tilde{Y}_1-1)$.
The BBN bound $\Omega_{\phi_1}<0.045$ amounts to 
\begin{equation}
\lambda_1 \gtrsim 9.42 \sqrt{2\tilde{Y}_1-1}\,.
\label{lamcon2}
\end{equation}
This shows that, under the condition $\tilde{Y}_1 \to \infty$, 
$\lambda_1$ needs to be infinitely large. 
However, as long as we do not demand the exact scaling 
radiation solution, the variable $Y_1$ can be of 
the order of unity (as we will see later). 
In such a case the constraint on $\lambda_1$ 
is not so restrictive. 

The radiation-dominated epoch can be followed by the 
scaling matter era characterized by the fixed point 
$(x_1, \tilde{Y}_1)=(\sqrt{6}/(2\lambda_1), 1)$ with 
$\Omega_{\phi_1}=3/\lambda_1^2$.
The solutions finally approach the assisted field-dominated
point satisfying Eq.~(\ref{gYre}), i.e. 
\begin{equation}
\tilde{Y} \equiv cY=\frac12+\frac{\lambda_{\rm eff}^2}{16}
\left( 1+\sqrt{1+\frac{16}{3\lambda_{\rm eff}^2}}
\right)\,,
\label{equy}
\end{equation}
where $c$ is the coefficient of the effective single-field 
Lagrangian density: $p=-1+cX$.
In deriving Eq.~(\ref{equy}) we have taken the solution with $\tilde{Y} \ge 1/2$.
{}From Eq.~(\ref{assiat}) the field equation of state is given by 
\begin{equation}
w_{\phi}=-1+\frac{\lambda_{\rm eff}}{2}
\left( \sqrt{\lambda_{\rm eff}^2+\frac{16}{3}}
-\lambda_{\rm eff} \right)\,,
\label{wphidi}
\end{equation}
which shows that the late-time cosmic acceleration 
occurs for $\lambda_{\rm eff}<\sqrt{6}/3$.
The stability of this solution is ensured for 
$\lambda_{\rm eff}^2<3(2\tilde{Y}-1)$, i.e. 
$\lambda_{\rm eff}<\sqrt{3}$.

\subsection{Two fields}

We first study cosmological dynamics of 
the two-field ghost condensate model.

Let us consider the case in which the field $\phi_1$ 
initially exists around $x_1 \simeq 2\sqrt{6}/(3\lambda_1)$
with a finite value of $\tilde{Y}_1$ satisfying 
the condition $\tilde{Y}_1 \ge 1/2$ [see Eq.~(\ref{lamxi})].
{}From Eq.~(\ref{auto4}) it follows that $\rd \tilde{Y}_1/\rd N>0$, 
as long as $x_1$ does not depart significantly from 
$2\sqrt{6}/(3\lambda_1)$.
This means that the quantity $\tilde{Y}_1$ tends to grow 
during most of the radiation era, which also leads to 
the increase of the density parameter 
$\Omega_{\phi_1}=x_1^2 (3 \tilde{Y}_1-1)$.
This growth of $\tilde{Y}_1$ is associated with the fact that 
the radiation scaling solution exists only in the limit 
$\tilde{Y}_1 \to \infty$.

\begin{figure}
\includegraphics[height=3.4in,width=3.4in]{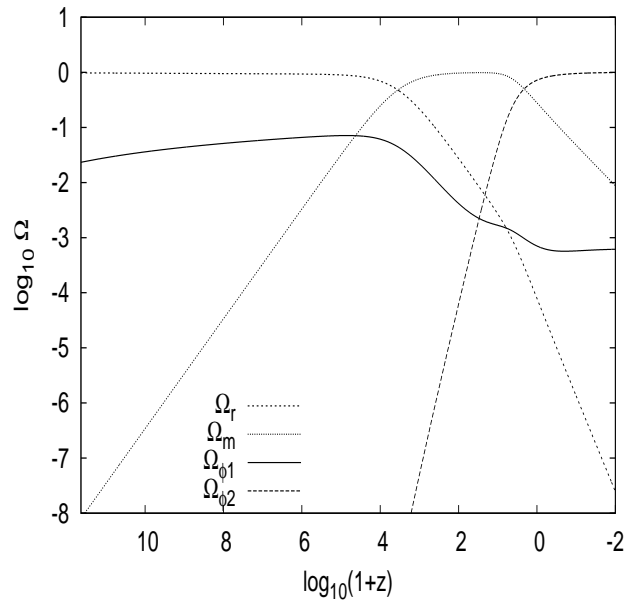}
\caption{\label{fig8}
Evolution of $\Omega_r$, $\Omega_m$, 
$\Omega_{\phi_1}$, and $\Omega_{\phi_2}$
versus the redshift $z$ for $\lambda_1=40$ and $\lambda_2=1$
in the two-field dilatonic ghost condensate model.
The initial conditions are chosen to be $x_1=2\sqrt{6}/(3\lambda_{1})$, 
$\tilde{Y}_1=5.0$, $x_2=1.6 \times 10^{-20}$, $\tilde{Y}_2=10.0$, 
and $\Omega_m=8 \times 10^{-9}$.
While the BBN bound $\Omega_{\phi_1}<0.045$ is satisfied 
at $z \approx 10^{10}$, the growth of $\Omega_{\phi_1}$ 
continues by the redshift at $z=7.2 \times 10^{4}$ with the maximum value
$\Omega_{\phi_1}=0.071$.}
\end{figure}
\begin{figure}
\includegraphics[height=3.5in,width=3.5in]{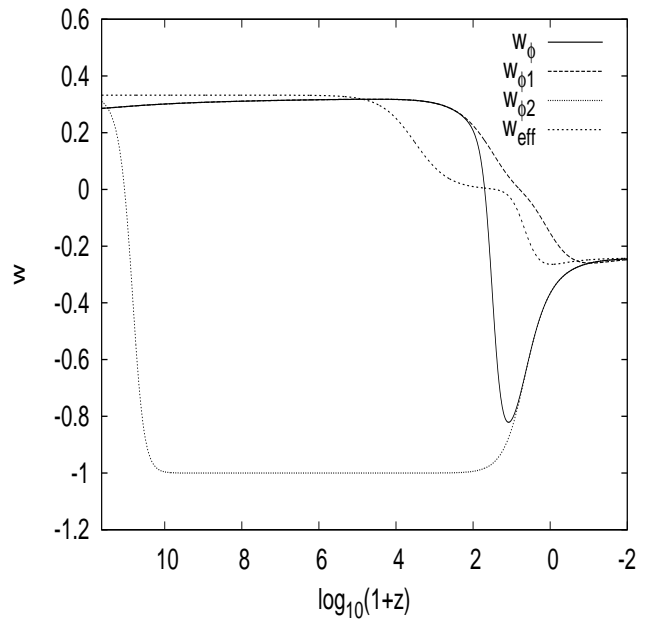}
\caption{\label{fig9}
Evolution of $w_{\phi}$, $w_{\phi_1}$, 
$w_{\phi_2}$, and $w_{\rm eff}$ for $\lambda_1=40$ 
and $\lambda_2=1$ in the two-field dilatonic 
ghost condensate model.
The initial conditions are the same as in Fig.~\ref{fig8}.}
\end{figure}

In Fig.~\ref{fig8} we plot one example about the evolution of 
density parameters for $\lambda_1=40$ and $\lambda_2=1$.
This shows that $\Omega_{\phi_1}$ in fact increases during 
the radiation-dominated epoch. 
In this case we have $\Omega_{\phi_1}=0.036$ around the BBN 
epoch ($z \approx 10^{10}$), so that the bound 
$\Omega_{\phi_1}<0.045$ is satisfied.
The growth of $\Omega_{\phi_1}$ ceases around the end
of the radiation era, because $\tilde{Y}_1$ begins to decrease
toward the scaling matter fixed point at $\tilde{Y}_1=1$.
In order to satisfy the BBN bound (\ref{lamcon2}) we have numerically found 
that $\lambda_1$ is required to be at least larger than 30.
This is related to the fact that, even for the initial conditions 
of $\tilde{Y}_1$ close to $1/2$ around the beginning of the radiation era,
$\tilde{Y}_1$ grows to be larger than 5 at the BBN epoch.

In Fig.~\ref{fig8} we find that the scaling matter era in which
$\Omega_{\phi_1}$ is nearly constant is very short, unlike 
the multi-field quintessence with exponential potentials.
This can be understood as follows.
Around the end of the radiation-dominated epoch 
the quantity $\tilde{Y}_1$ has already increased to 
a value larger than the order of unity.
It takes some time for the solutions to reach the scaling matter
fixed point at $\tilde{Y}_1=1$. 
In the numerical simulation of Fig.~\ref{fig8} this happens for 
the redshift at $6.2$. Since the solutions enter the dark energy 
dominated epoch for $z \lesssim {\cal O}(1)$, the period of 
the scaling matter era is short.

Figure \ref{fig9} illustrates the evolution of $w_{\phi}$, $w_{\phi_1}$, 
$w_{\phi_2}$, and $w_{\rm eff}$ for the same model parameters 
and initial conditions as given in Fig.~\ref{fig8}.
Initially $w_{\phi_1}$ is smaller than $w_{\rm eff}$, but it grows
to the value close to $w_{\rm eff}=1/3$ during the radiation era 
with the increase of $Y_1$. 
The evolution of $w_{\phi_1}$ in Fig.~\ref{fig9} clearly shows that 
the field $\phi_1$ does not soon enter the scaling matter regime
just after the radiation-dominated epoch.
The field $\phi_2$ approaches the phase with $w_{\phi_2} \simeq -1$ 
after the initial transient period. This corresponds to $\tilde{Y}_2 \simeq 1/2$
and $\Omega_{\phi_2} \simeq x_2^2/2$ in Eq.~(\ref{wphigho}). 
Numerically we find that the late-time cosmological evolution is practically 
independent of the initial conditions of $Y_2$, but it is sensitive to the 
initial values of $x_2$ because the quantity $x_2$ is associated 
with the dark energy density.

\begin{figure}
\includegraphics[height=3.2in,width=3.5in]{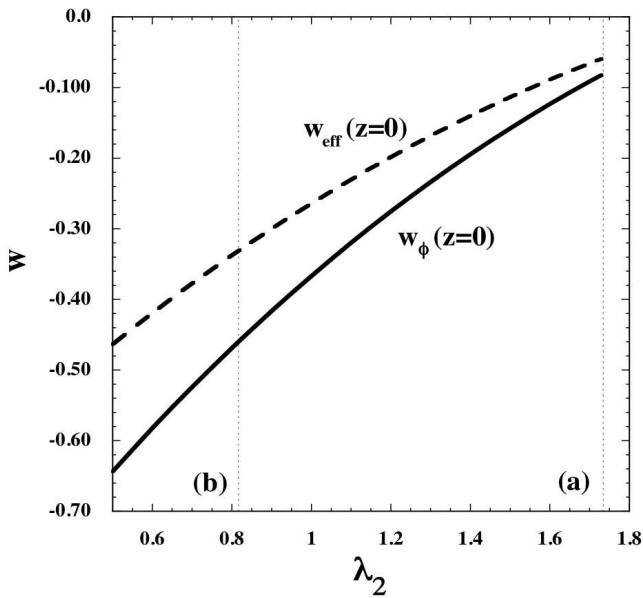}
\caption{\label{fig10}
The equations of state $w_{\phi}$ and $w_{\rm eff}$ today 
versus $\lambda_2$ for $\lambda_1=40$
in the two-field dilatonic ghost condensate model.
If $\lambda_2<1.734$ the scalar-field dominated 
point is the final attractor. 
If $\lambda_2>\sqrt{6}/3$ the field $\phi_2$ cannot be 
responsible for cosmic acceleration as a single 
component of the universe.
The two lines (a) and (b) in the figure correspond to $\lambda_2=1.734$
and $\lambda_2=\sqrt{6}/3$, respectively.}
\end{figure}

The epoch at which the field $\phi_2$ starts to exit from the regime 
$w_{\phi_2}=-1$ depends on the parameter $\lambda_2$. 
For decreasing $\lambda_2$ the redshift $z_c$ at which this
``thawing'' occurs gets smaller, which leads to smaller values of
$w_{\phi}$ and $w_{\rm eff}$ today.
In Fig.~\ref{fig10} we plot $w_{\phi}(z=0)$ and $w_{\rm eff}(z=0)$
versus $\lambda_2$ for $\lambda_1=40$.
In this case the stability of the assisted field-dominated point 
is ensured for $\lambda_2<1.734$. 
If $\lambda_2>\sqrt{6}/3$ the field $\phi_2$ cannot drive 
cosmic acceleration as a single component of the universe.
Under these bounds we find that the condition $w_{\rm eff}(z=0)<-1/3$
for the acceleration today is not satisfied in the two-field case.
This is intimately associated with the fact that the ``thawing'' of the 
field $\phi_2$ occurs quite early ($z_c >{\cal O}(10)$) 
for $\lambda_2>\sqrt{6}/3$, see Fig.~\ref{fig9}.
This property is different from two-field quintessence 
with exponential potentials in which $\phi_2$ begins to evolve
at smaller redshifts for the same values of $\lambda_2$.
We also note that the present values of $w_{\phi}$ and $w_{\rm eff}$
are insensitive to the choice of $\lambda_1$, 
as long as the condition (\ref{lamcon2}) is satisfied.

\subsection{More than two fields}
\begin{figure}
\includegraphics[height=3.2in,width=3.5in]{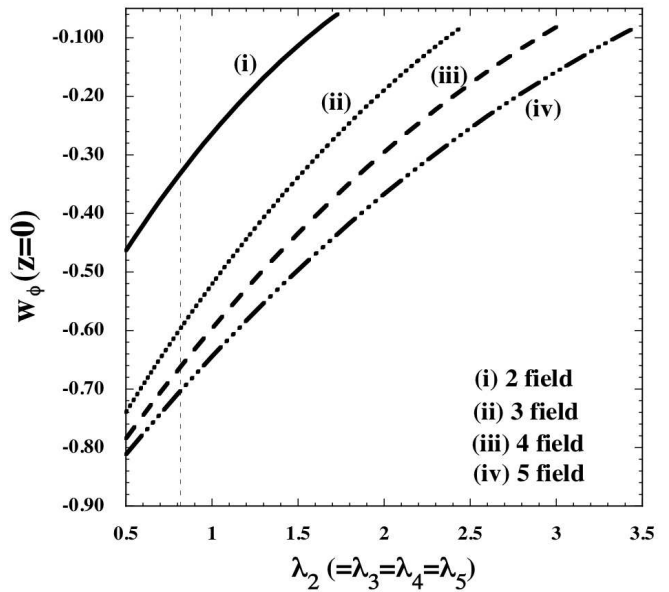}
\caption{\label{fig11}
The equation of state $w_{\phi}$ today 
versus $\lambda_i$ ($i \neq 1$) for $\lambda_1=40$
in the multi-field dilatonic ghost condensate model 
(up to five fields with the same slopes $\lambda_i$, $i \ge 2$).
In this simulation the fields $\phi_i$ ($i \ge 2$) enter 
the regime of the assisted attractor almost at the same time. 
Under the condition $\lambda_i>\sqrt{6}/3$ (which
is shown as a thin dotted line in the figure), 
it is difficult to realize $w_{\phi} (z=0)<-0.8$ even 
in the presence of five scalar fields.}
\end{figure}
\begin{figure}
\includegraphics[height=3.2in,width=3.5in]{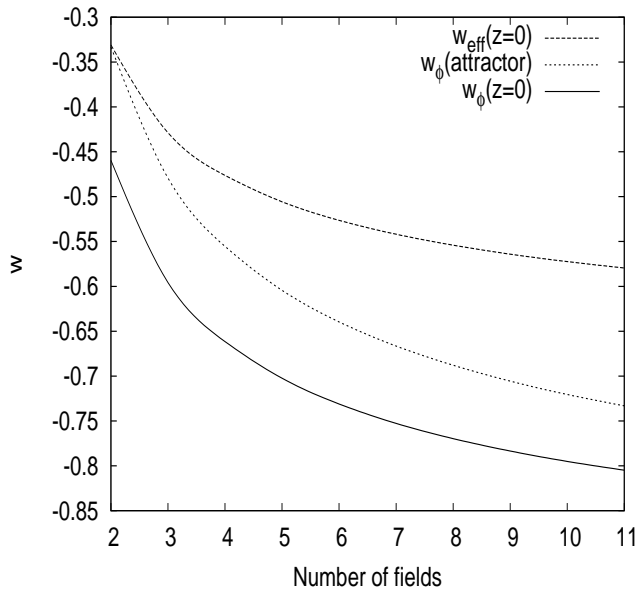}
\caption{\label{fig12}
The present equations of state $w_{\phi}(z=0)$ and 
$w_{\rm eff}(z=0)$ and $w_{\phi}$ at the assisted attractor  
versus the number $n$ of scalar fields for $\lambda_1=40$
and $\lambda_i=0.817$ ($i \ge 2$) in the multi-field
dilatonic ghost condensate model.
In this simulation the fields $\phi_i$ ($i \ge 2$) enter 
the regime of the assisted attractor almost at the same time.}
\end{figure}

In the presence of more than two scalar fields it is possible to obtain 
smaller values of $w_{\phi}$ and $w_{\rm eff}$ today relative to 
the two-field scenario discussed above.
In Fig.~\ref{fig11} we plot $w_{\phi}(z=0)$ versus $\lambda_i$ ($i \ge 2$)
for $\lambda_1=40$. The initial conditions for $\phi_i~(i \ge 2)$
are chosen so that they join the assisted attractor almost at the same time.
For the three-field scenario with $\lambda_2=\lambda_3=0.817$
(slightly larger than $\sqrt{6}/3$) one has $w_{\phi}(z=0)=-0.60$ and 
$w_{\rm eff}(z=0)=-0.43$, 
so that cosmic acceleration is realized today.
However this case is difficult to be compatible 
with the observational bound $w_{\phi}(z=0)<-0.8$.
Even for the five-field case with $\lambda_i=0.817$ ($i=2, 3, 4, 5$)
we find that $w_{\phi}(z=0)=-0.70$, which is still larger than $-0.8$.

In Fig.~\ref{fig12} we plot $w_{\phi}(z=0)$, $w_{\rm eff}(z=0)$, and
$w_{\phi}$ at the late-time attractor for $\lambda_1=40$ and 
$\lambda_i=0.817$ ($i \ge 2$).
This shows that we require at least 10 scalar fields to realize
the condition $w_{\phi} (z=0)<-0.8$.
Equation (\ref{wphidi}) leads to larger $w_{\phi}$ 
at the assisted attractor relative to the case of the multi-field 
quintessence with exponential potentials for the same 
values of $\lambda_{\rm eff}$.
In addition to the early ``thawing'' of assisting scalar fields, 
this is another reason why a large number of fields are required
to realize small $w_{\phi}(z=0)$ close to $-1$.
In Fig.~\ref{fig12} we find that $w_{\phi}(z=0)$ is almost 
proportional to $w_{\phi}({\rm attractor})$ for $n \ge 3$.
Unless we have many fields such that $n \ge 10$, 
$w_{\phi}(z=0)$ as well as $w_{\phi}({\rm attractor})$
are not reduced sufficiently to satisfy the observational bound.
Of course, if we do not demand the condition $\lambda_i>\sqrt{6}/3$ 
($i \ge 2$), it is possible to realize $w_{\phi}(z=0)<-0.8$ without 
introducing many fields.

\section{Conclusions}
\label{conclusions}

In this paper we have studied cosmological dynamics of assisted dark energy 
for the Lagrangian density (\ref{psum}) that possesses scaling solutions.
This scaling Lagrangian density involves many models such as quintessence 
with exponential potentials, dilatonic ghost condensates, and tachyon fields
with inverse power-law potentials.
As long as the energy density of a field $\phi_1$ (with $\lambda_1^2 \gg p_{,X_1}$) 
dominates over those of other fields, the density parameter $\Omega_{\phi_1}$ 
remains constant during the radiation and matter eras 
($\Omega_{\phi_1}=4p_{,X_1}/\lambda_1^2$ and 
$\Omega_{\phi_1}=3p_{,X_1}/\lambda_1^2$, respectively).
This property is attractive because the solutions enter the scaling regime 
even if the field energy density is initially comparable to 
the background fluid density.

In the presence of multiple scalar fields the scaling matter era
can be followed by the phase of a late-time cosmic acceleration
as long as more than one field join the assisted attractor. 
The field equation of state for the assisted attractor takes 
an effective single-field value 
$w_{\phi}=-1+\lambda_{\rm eff}^2/(3p_{,X})$,
with $\lambda_{\rm eff}$ given by Eq.~(\ref{lameff}).
Since $\lambda_{\rm eff}$ is smaller than the slope $\lambda_i$
of the each field, the presence of multiple scalar fields can 
give rise to cosmic acceleration even if none is able to 
do so individually. This is a nice feature from the viewpoint 
of particle physics because there are in general many scalar
fields (dilaton, modulus, etc) with the slopes $\lambda_i$
larger than the order of unity.

While the above property of cosmological dynamics is generic 
for the scaling models with the Lagrangian density (\ref{psum}), 
the evolution of $w_{\phi}$ as well as $\Omega_{\phi}$
is different depending on the forms of the Lagrangian 
density $p$. In order to see this we have focused on two models: 
(i) canonical fields with exponential potentials, and 
(ii) multiple dilatonic ghost condensates.
These correspond to representative examples of 
quintessence and k-essence, respectively.

For the multi-field quintessence with exponential potentials,
the slope $\lambda_1$ for the scaling field is constrained to be 
$\lambda_1>9.42$ from the BBN bound.
We have numerically found that the transient cosmic acceleration 
today with a non-accelerated attractor can be realized after the 
scaling matter era. This comes from the thawing property of 
assisting scalar fields that start to evolve only recently from 
a nearly frozen regime characterized by the equation of state 
$w_{\phi_i} \simeq -1$ ($i \ge 2$).
Even for the initial conditions where the kinetic energies 
of $\phi_i$ ($i \ge 2$) are much larger than their potential 
energies, we have confirmed that the fields $\phi_i$ enter
the frozen regime by the end of the 
radiation-dominated epoch.
In the presence of three scalar fields we have found that the 
total field equation of state $w_{\phi}$ today can be smaller 
than $-0.8$, even if each field is unable to be responsible for 
the accelerated expansion as a single component of the universe.

The multi-field dilatonic ghost condensate model does not possess an exact 
scaling radiation era, although the scaling matter era is present.
In this model the slope $\lambda_1$ of the field $\phi_1$ is 
more severely constrained from the BBN bound relative to the
multi-field quintessence with exponential potentials. 
The fields $\phi_i$ ($i \ge 2$) enter the regime characterized
by $w_{\phi_i} \simeq -1$ during the radiation-dominated epoch.
However the exit from this regime occurs earlier than the 
multi-field quintessence with exponential potentials
for the same values of $\lambda_i$, 
which generally leads to a larger field equation of state $w_{\phi}$ today.
In the two-field scenario we have found that cosmic acceleration 
does not occur at the present epoch if the fields are unable to give rise
to inflation individually. While the acceleration today is possible
in the presence of more than two fields, we require at least 10
fields to satisfy the observational bound $w_{\phi}(z=0)<-0.8$
under the condition $\lambda_i>\sqrt{6}/3$ ($i \ge 2$).

In single-field scaling models with an exit to the late-time acceleration
(such as the model in Ref.~\cite{Barreiro}),
the field equation of state $w_{\phi}$ changes from 0 to negative
during the transition from the matter era to the accelerated epoch.
Meanwhile the multi-field dark energy models we have discussed
in this paper exhibit a rather peculiar behavior of $w_{\phi}$:
it first reaches a minimum and then starts to grow toward 
the assisted attractor (see Figs.~\ref{fig4} and \ref{fig9}).
It will be of interest to see whether future high-precision 
observations will detect some signatures for such dynamics.

\section*{ACKNOWLEDGEMENTS}
ST thanks Reza Tavakol for very kind hospitality during his stay in 
Queen Mary, University of London, where this work was 
partially done.
We thank Hitoshi Fujiwara for useful discussions.
ST thanks financial support for JSPS (No.~30318802).

\end{document}